\documentclass{ws-procs9x6}            % default, citations in superscript
\begin{document}
\title{Entanglement Entropy for time dependent  two dimensional holographic superconductor }

\author{N.S. Mazhari$^{*,1}$, D. Momeni $^{*,2}$, Kairat Myrzakulov $^{*,3}$, R. Myrzakulov$^{*,4}$}

\address{$^*$ Department of General and Theoretical Physics, Eurasian national University,\\
Astana, 010008, Kazakhstan\\
$^1$E-mail: najmemazhari86@gmail.com\\$^2$E-mail:momeni\_d@enu.kz\\$^3$E-mail:myrzakulov\_kr@enu.kz\\$^4$E-mail:rmyrzakulov@gmail.com\\
www.enu.kz}

\begin{abstract}
We studied entanglement entropy for a time dependent two dimensional holographic superconductor. We showed that the conserved charge of the system plays the role of the critical parameter to have condensation. 
\end{abstract}


\bodymatter
%%%%%%%%%%%%%%%%%%%%%%%
\section{Introduction}
Entanglement  exists in the core of many aspects of quantum field theory and statistical mechnaics and it is therefore desirable to understand its structure as well as possible using gauge-grvaity duality \cite{CY}-\cite{AJ} . One attempt to improve our understanding of entanglement is the study of our ability to perform calculations   on asymptotic Anti de Sitter bulk models. This is due to the celebrated anti de Sitter/conformal field theory (AdS/CFT) conjecture, which it relates the weakly coupled gravitational theory in bulk to the strongly coupled quantum theory in a flat boundary \cite{Maldacena}. This conjecture provides a framework to study the role of entanglement entropy of a quantum system in terms of minimal surfaces \cite{hee1,hee2}.

 In \cite{hee1,hee2} it was shown that the holographic entanglement entropy (HEE) of a quantum system in boundary is defined as the entropy of a region of space $\tilde{A}$ and its complement  on the minimal surfaces in $AdS_{d+1}$ using gauge-gravity duality \cite{hee1},\cite{hee2}:

\begin{eqnarray}
S_{\tilde{A}}\equiv S_{HEE}=\frac{Area(\gamma _{\tilde{A}})}{4G_{d+1}}.\label{HEE}
\end{eqnarray}%
For time-independent backgrounds we need to compute the minimal area of a region in bulk with the same boundary $\partial A$ with the quantum system in boundary. 

 Another surprising development is that there exist a time dependent formulation of HEE  using extremal surfaces instead of minimal ones \cite{Hubeny:2007xt}. THis approach  used to study HEE in the  time-dependent Janus background \cite{Ugajin:2014nca}. The technique is to replace "minimal" with "extremal" surfaces:
\begin{eqnarray}
S^{time-dependent}_{\tilde{A}}\equiv S_{HEE}=ext\Big[\frac{Area(\gamma _{\tilde{A}})}{4G_{d+1}}\Big].\label{HEE}
\end{eqnarray}
In case of multiple extremal surfaces, we should select the extremal surface with the minimum area included in them.\par
 Here we discuss the issue of HEE for two dimensional time dependent holographic superconductors  \cite{Liu:2011fy}-\cite{Momeni:2015iea} , and show that in this regime the HEE and phase transition can be achieved.

%%%%%%%%%%%%%%%%%%%%%%%%%%%%%%%%%%
\section{Phase diagram of $2D$ holographic superconductors }\label{Far-from}

The bulk action with  an Abelian gauge field $A$ in the presence of a generally massive scalar field $\Psi$ with charge $q$ and mass $m$ was selected in order to determine the two dimensional holographic superconductor
 \cite{HHH1}-\cite{Hartnoll:2008vx},
\begin{eqnarray}\label{S}
&&S=
\int_\mathcal{M} d^3x\sqrt{-g}[\frac{1}{2\kappa^2}\Big(R+\frac{2}{l^2}\Big)+(-\frac{1}{4}F_{ab}F^{ab}-|D\Psi|^2-m^2|\Psi|^2)],
\end{eqnarray}
here
 $l$ is the AdS radius, and $D=\nabla-iA$ with $\nabla\equiv \partial_{\mu}$, Maxwell tensor is $F_{\mu\nu}=\partial_{\mu}A_{\nu}-\partial_{\nu}A_{\mu}$. For the sake of simplicity, only a probe limit $\kappa^2=0,l=1,q=1,z_h=1$  was analysed .   
The set of equations of motion (EOMs) for action Eq. (\ref{S}) derived from the basic variation with respect to the scalar field and gauge fields as follo:
 \begin{eqnarray}
 &&D_\mu D^\mu\Psi-m^2\Psi=0,\\
&& \nabla_\mu F^{\mu\nu}=i[\Psi^* D^\nu\Psi-\Psi (D^\nu\Psi)^*].
 \end{eqnarray}
The advantage of using AdS/CFT set up was that the dual CFT temperature could be obtained as the Hawking temperature of the horizon (denoted by $z_h$) by the following formula:
\begin{equation}
T=\frac{1}{2\pi z_h}\label{T}.
\end{equation}
The  AdS boundary was located at $z=0$, which provided initial data for scalar field and  of
 a  physical sources for the scalar operator $\mathcal{O}$. Because the conformal scaling was used for field theory, conformal  dimension was defined
using two conformal dimensions 
 $\Delta_{\pm}=1\pm\sqrt{1+m^2l^2}$. In this report we study the case of $m^2l^2 =0$,$\Delta=\Delta_{+}=2$\\
%%%%%%%%%%%%%%%%%%%%%%%%%%%%%%%%%%%%%%%%
The \emph{ vacuum expectation value (VEV) of the corresponding boundary quantum field theory operators} basically can be computeted using the following variations of the effective and renormalized  action $\delta S_{ren}$:
 \begin{eqnarray}
&& \langle J^\nu\rangle=\frac{\delta S_{ren}}{\delta a_\nu}=\lim_{z\rightarrow 0}\frac{\sqrt{-g}}{q^2}F^{z\nu},\\
&&\langle O\rangle=\frac{\delta S_{ren}}{\delta\phi}=\lim_{z\rightarrow 0}[\frac{z\sqrt{-g}}{lq^2}(D^z\Psi)^*-\frac{z\sqrt{-\gamma}}{l^2q^2}\Psi^*],
 \end{eqnarray}
 where $S_{ren}$ is the renormalized action:
 \begin{equation}
 S_{ren}=S-\frac{1}{lq^2}\int_\mathcal{B}\sqrt{-\gamma}|\Psi|^2,
 \end{equation}
and by dot we mean derivative with respect to the time coordinate  $t$. 
\par
%%%%%%%%%%%%%%%%%%%%%%%%%%%%%%%%%%%
\par
\section{Calculation of holographic entanglement entropy }\label{Calculation of holographic}

An alternative form for metric could be find using the coordinate transformation  $y=t+2\tanh^{-1} z$ in  the metric on AdS boundary. Under this transformation the metric becomes flat which is neccesary to construct CFT. The new non static (time dependent) form of metric is given by the following:
\begin{eqnarray}
&&
ds^2=\frac{l^2}{\tanh^2\Big(\frac{t-y}{2}\Big)}\Big[dx^2-\cosh^{-2}\Big(\frac{t-y}{2}\Big)dtdy\Big]\label{metric2}.
\end{eqnarray}
 In this coordinate $y$, the black hole horizon $z_h=1$ mapps to $y = \infty$ and  the conformal (AdS) boundary $z=0$ mapps to the $y-t=0$. Note that, so far  the metric on conformal boundary is manifested as flat in the form $ds^2\sim dx^2-dt^2$.  We now calculate entanglement entropy given by Eq. (\ref{HEE}) \par
%%%%%%%%%%%%%%%%%%%%%%%%%
\subsection{Extremal areas in connected phase}
For connected surfaces, we should regularized HEE per length $l$ as a function of the $\{u(y_{\infty}),c\}$.  It was shown that the area per entangled length $\frac{A}{l}$ is a
monotonic-increasing function. This quantity shows that the system remains on a regular phase of
matter for $T>T_c$ or equivalently for $u(y_{\infty})\succeq15.5$. Regular attendance at these non superconducting
phase has proved numerically. Boundary conditions and regular tiny
parameter $c$ will help to keep normal phase for longer. A more interesting observation is that the 
hardenability to form normal phase becomes more harder with the increase of entropy $\frac{A}{l}$.

%%%%%%%%%%%%%%%%%%%%%%%
\subsection{ Extremal areas in disconnected phase}
In disconnected phase, we should compute the extremal area of the disconnected surfaces as a function of boundary coordinates $(t_{\infty},x_{\infty})$.We show that the coserved charge $J$ of the associated Euler-Lagrange system of the Eqs. plays the role of the critical parameter. 
A numerical computation shows that when $J=\frac{1}{2}$, the normalized area per length  $\frac{A}{l}$ is a
monotonic-decreasing function.  Near $J_c\approx 0.58^{+}_{-}0.02$ we can simplify write the time evolution of the extremal path as the following  $\dot{u}=T_1+T_2(J-0.5)$ where $T_i$ are functions of $u$. In this case with disconnected areas, we detect that the extremal area per length 
 is always decreasing. It produces a regular phase of
matter for $J>J_c$. 
%%%%%%%%%%%%%%%%%%%%%%
Further more we can show that the HEE is  a linear function of total length $y_{\infty}$. The simple physical reason backs to the \emph{emergence of new extra degrees of freedom
in small values of belt length (small sizes)}.

%%%%%%%%%%%%%%%%%%%%%%%%%%%%%%%%%%%

\par
\section{Summary}\label{Summary}
We investigated holographic entanglement entropy of dual quantum systems using a time dependent version of AdS spacetime.  By a numerical computation we demonstrated that  HEE is a
monotonic-decreasing function.  It is always decreasing. It produces a regular phase of
matter for a critical value $J>J_c$. 
  We showed that $\log(S_{HEE})$ is proportional to the numbers of degrees of freedom (dof) of the system $\mathcal{N}$. Just as the phenomena of sudden condensate,  we observed an emergence of new dof in system .
 The HEE growing and exhibiting its activity in accordance with thermodynamic laws.
Furthermore we demonstrate that the HEE as function of $J$ and $u(y_{\infty})$ when time $y$ is growing up and for  large values of $J$ reaches a local maxima. So, system undergoes a normal thermodynamically time evolution according to the second law. Consequently the arrow of time never changes.

%\section{Acknowledgments}

\end{document}